\documentclass[12pt]{iopart}
\usepackage{iopams}
\usepackage{amsfonts}
\usepackage{amsmath}
\usepackage{amssymb}
\usepackage{mathabx}
\usepackage{mathrsfs}
\usepackage{array}
\usepackage{tensor}
\usepackage{esint}
\usepackage{tikz}
\usepackage{harvard}
\usepackage{hyperref}
\usepackage{cite}
\usepackage{algorithmic}
\usepackage{graphicx}
\usepackage{textcomp}
\usepackage{xcolor}
\usepackage{multicol,multirow}

\begin{document}
\title[]{Machine Learning-based Signal Quality Assessment for Cardiac Volume Monitoring in Electrical Impedance Tomography} 
\author{Chang Min Hyun$^{1}$, Tae Jun Jang$^{1}$, Jeongchan Nam$^{2}$, Hyeuknam Kwon$^{3}$, Kiwan Jeon$^{4}$, and Kyunghun Lee$^{5}$\footnote[5]{To whom correspondence should be addressed (imlkh84@gmail.com)}}
\address{$^{1}$School of Mathematics and Computing (Computational Science and Engineering), Yonsei University, Seoul, Republic of Korea.}
\address{$^{2}$BiLab, Pangyo, Republic of Korea.}
\address{$^{3}$Division of Software, Yonsei University, Wonju, Republic of Korea.}
\address{$^{4}$National Institute for Mathematical Sciences, Daejeon, Republic of Korea.}
\address{$^{5}$Kyung Hee University, Seoul, Republic of Korea.}

\begin{abstract}
Owing to recent advances in thoracic electrical impedance tomography, a patient's hemodynamic function can be noninvasively and continuously estimated in real-time by surveilling a cardiac volume signal associated with stroke volume and cardiac output. In clinical applications, however, a cardiac volume signal is often of low quality, mainly because of the patient's deliberate movements or inevitable motions during clinical interventions. This study aims to develop a signal quality indexing method that assesses the influence of motion artifacts on transient cardiac volume signals. The assessment is performed on each cardiac cycle to take advantage of the periodicity and regularity in cardiac volume changes. Time intervals are identified using the synchronized electrocardiography system. We apply divergent machine-learning methods, which can be sorted into discriminative-model and manifold-learning approaches. The use of machine-learning could be suitable for our real-time monitoring application that requires fast inference and automation as well as high accuracy. In the clinical environment, the proposed method can be utilized to provide immediate warnings so that clinicians can minimize confusion regarding patients' conditions, reduce clinical resource utilization, and improve the confidence level of the monitoring system. Numerous experiments using actual EIT data validate the capability of cardiac volume signals degraded by motion artifacts to be accurately and automatically assessed in real-time by machine learning. The best model achieved an accuracy of 0.95, positive and negative predictive values of 0.96 and 0.86, sensitivity of 0.98, specificity of 0.77, and AUC of 0.96.
\end{abstract}

\maketitle

\section{Introduction}
\begin{figure}[h]
	\centering
	\includegraphics[width=0.65\textwidth]{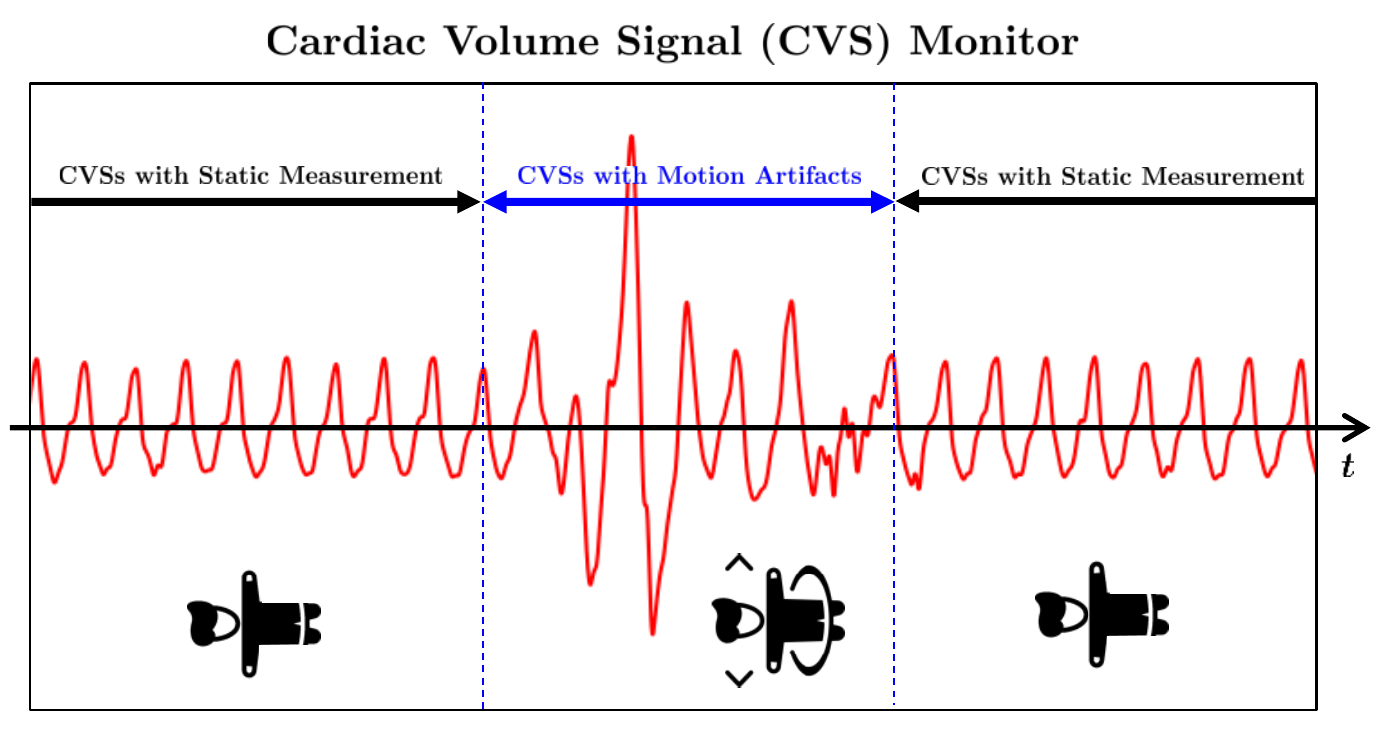}
	\caption{Motion-induced artifacts in cardiac volume monitoring using electrical impedance tomography. Patient's deliberate movements or inevitable motions during clinical intervention cause severe artifacts in a cardiac volume signal.}
	\label{CVSmotionartifacts}
\end{figure}

Over several decades, continued advances in electrical impedance tomography (EIT) have expanded the clinical capability of real-time cardiopulmonary monitoring systems by overcoming the limitations of traditional methods, such as cardiac catheterization through blood vessels \cite{Adler2017,Borges2012,Deibele2008,Frerichs2014,Kubicek1970,Kerrouche2001,Lee2018,Putensen2019,Zlochiver2006}. Recently, based on thoracic EIT, a patient's hemodynamic function can be noninvasively and continuously estimated in real-time by surveilling a signal extracted using EIT, the so-called cardiac volume signal (CVS), which has a strong relationship with key hemodynamic factors such as stroke volume and cardiac output \cite{Jang2020,Askari2019,Westterhof2019}. In clinical applications, however, a cardiac volume signal is often of low quality, mainly because of the patient's deliberate movements or inevitable motions during clinical interventions such as medical treatment and nursing. Because postural change causes movement of the chest boundary to which existing EIT solvers are highly sensitive owing to time-difference-reconstruction characteristics \cite{Adler1996,Boyle2010,Lionheart1998,Lee2017,Seo2013}, motion-induced artifacts are generated in the CVS, as shown in Figure \ref{CVSmotionartifacts}. 

CVS extraction is to separate a cardiogenic component from the EIT voltage data, resulting from current injections at electrodes attached across a human chest. In recent studies \cite{Jang2020,Lee2021}, effective CVS extraction was successful in motion-free measurements where voltage data are mainly influenced by air and blood volume changes in the lungs, heart, and blood vessels comprehensively, but not by motions. In contrast, achieving the cardiogenic component separation in motion-influenced measurements is still a long-term challenge. Postural changes in EIT measurements cause strong distortion of the voltage data \cite{Adler1996,Zhang2005} and easily disturb the extraction of relatively weak cardiogenic signals \cite{Brown1992,Pikkemaat2014,Leonhardt2012}.

Handling motion interference has been a huge challenge in most EIT-based techniques for enhancing clinical capability, but not researched much yet \cite{Yang2022}. Adler \textit{et al.} \cite{Adler1996} and Zhang \textit{et al.} \cite{Zhang2005} investigated the negative motion effect in the EIT. Soleimani \textit{et al.} \cite{Soleimani2006} and Dai \textit{et al.} \cite{Dai2008} proposed a motion-induced artifact reduction method by reconstructing electrode movements along with conductivity changes. Lee \textit{et al.} \cite{Lee2017} analyzed motion artifacts in EIT measurements and proposed a subspace-based artifact rejection method. Yang \textit{et al.} \cite{Yang2022} suggested the discrete wavelet transform-based approach that reduces motion artifacts of three specific types. However, clinical motion artifacts are still not effectively addressed because of practical motion's immense diversity and complexity. Accordingly, for the time being, the EIT-based hemodynamic monitoring system attempts to be preferentially developed toward filtering motion-influenced CVSs rather than recovering them. In the clinical environment, this filtration can provide immediate warnings so that clinicians can minimize confusion regarding the patient's condition, reduce clinical resource utilization, and improve the confidence level of the monitoring system \cite{Charlton2021}. 

This study aims to develop a signal quality indexing (SQI) method that assesses whether motion artifacts influence transient CVSs. To take advantage of the periodicity and regularity in cardiac volume changes, the assessment is performed on each cardiac cycle, whose time intervals are identified using the synchronized electrocardiography (ECG) system. We leverage machine learning (ML), which has provided effective solutions for various biosignal-related tasks through feature disentanglement of complicated signals \cite{Alfaras2019,Belo2017,Celin2018,Hyun2021,Lecun2015,Seo2019,Sahoo2020,Wasimuddin2020}. The use of ML could be suitable for our real-time monitoring application that requires fast inference and automation as well as high accuracy.

We apply divergent ML methods, which can be sorted into discriminative-model and manifold-learning approaches. The discriminative-model approach is first considered, where an SQI map is directly trained using a paired dataset of CVS and its label \cite{Cramer2002,Hinton2007,Simonyan2014}. Although this approach provides a high performance on a fixed dataset, owing to the class imbalance problem, there is a risk of overfitting on motion-influenced CVS data in the scope of generalization or stability \cite{Buda2018,Cao2019,He2009,Van2017}. Motion artifacts can vary considerably in real circumstances, whereas collecting CVS data in numerous motion-influenced cases is practically limited because of the high cost, intensive labor, security, and ambiguity in clinical data acquisition and annotation \cite{Chapelle2009,Schlegl2017,Van2020,Zhu2009}. To handle this conceivable difficulty, the manifold-learning approach \cite{Jolliffe2016,Kingma2013,An2015,Higgins2016} is examined as an alternative. It does not learn irregular and capricious patterns of motion-influenced CVSs and only takes advantage of the learned features from motion-free CVSs.

Numerous experiments have been conducted using actual EIT data. Empirical results demonstrate that discriminative and manifold-learning models provide accurate and automatic detection of motion-influenced CVS in real-time. The best discriminative model achieved an accuracy of 0.95, positive and negative predictive values of 0.96 and 0.86, sensitivity of 0.98, specificity of 0.77, and AUC of 0.96. The best manifold-learning model achieved accuracy of 0.93, positive and negative predictive values of 0.97 and 0.71, sensitivity of 0.95, specificity of 0.80, and AUC of 0.95. The discriminative models yielded a more powerful SQI performance; in contrast, the manifold-learning models provided stable outcomes between the training and test sets. Regarding to practical applications, the choice of two models relies on what should be emphasized in the monitoring system in terms of performance and stability.

\section{Methods}
\begin{figure}[h]
	\centering
	\includegraphics[width=1\textwidth]{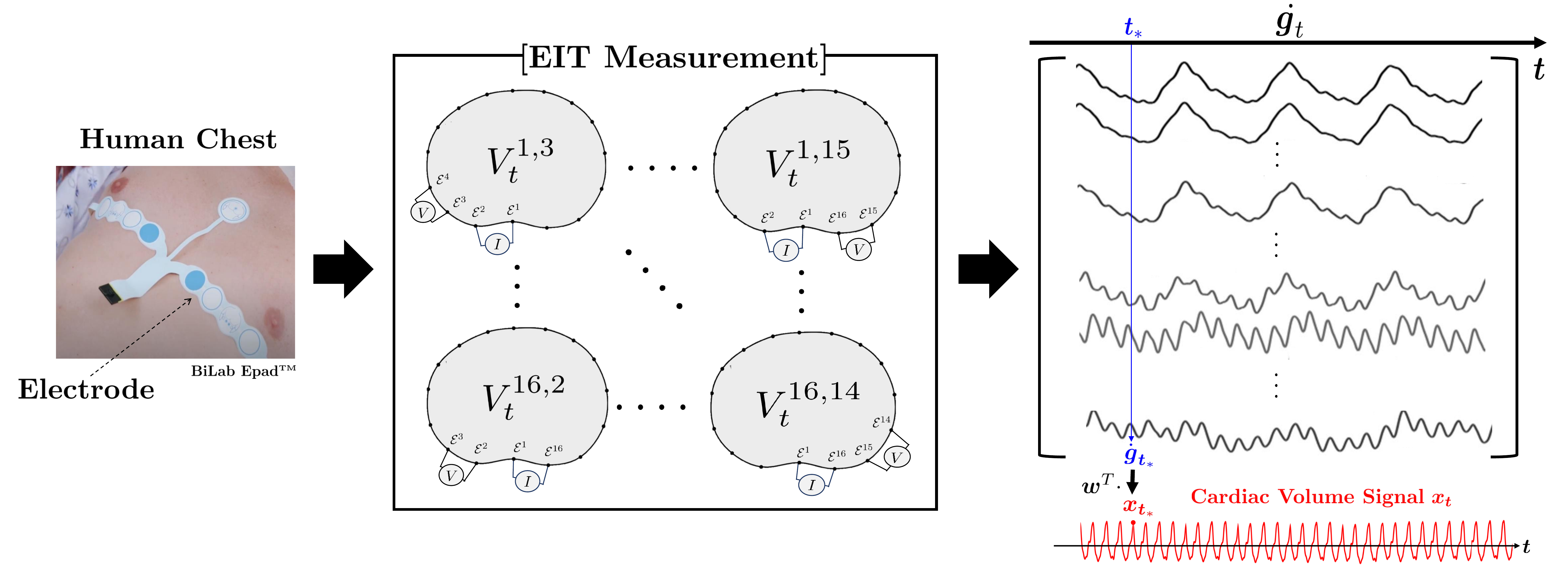}
	\caption{16-channel system of thoracic EIT and CVS extraction. The EIT machine measures voltage differences by injecting currents via electrodes attached along human chest. A cardiac volume signal $\boldsymbol x_{\boldsymbol t}$ is extracted by taking suitable weighting $\boldsymbol w$ to the time-difference transconductance $\dot{\boldsymbol g}_t$, which is defined by measured voltage data. Here, $\boldsymbol w$ is called as a leadforming vector, which is designed to separate a cardiogenic trans-conductance change from superposed data $\dot{\boldsymbol g}_t$ \cite{Lee2021}.}
	\label{EItsystem}
\end{figure}

This study considers the 16-channel system of the thoracic EIT, where 16 electrodes are attached along the human chest (see Figure \ref{EItsystem}). The EIT system is assumed to be synchronized with the ECG system, which provides the time interval for each cardiac cycle. The EIT device measures a set of voltage differences by injecting an alternative current of $I$ (mA) through pairs of adjacent electrodes while keeping all other electrodes insulated. At sampling time $t$, the following voltages are acquired: 
\begin{equation} \label{voltagedata}
	\{ V^{j,k}_{t} ~ : ~ V^{j,k}_t =  U_t^{j,k} - U_t^{j,k+1}, j \in \mathcal I, k \in \mathcal I \backslash \{j,j+1\} \} 
\end{equation} 
where $\mathcal I$ is an index set defined by $\mathcal I = \{1,2,\cdots,16\}$, $\mathcal E_k$ is the $k$-th electrode, and $U_t^{j,k}$ is the electrical potential on $\mathcal E_k$ subject to the current injection from $\mathcal E_j$ to $\mathcal E_{j+1}$. For notational convenience, $\mathcal E^{0}$ and $\mathcal E^{17}$ can be understood as $\mathcal E^{16}$ and $\mathcal E^{1}$, respectively. Once the current is injected from $\mathcal E^{j}$ to $\mathcal E^{j+1}$ for some $j \in \mathcal I$, the voltage is measured at each of the 16 adjacent electrode pairs $(\mathcal E^k, \mathcal E^{k+1})_{k \in \mathcal I}$. Among the 16 voltages, $V_t^{j,j-1}$, $V_t^{j,j}$, and $V_t^{j,j+1}$ are discarded to reduce the influence of the skin-electrode contact impedance \cite{Seo2013}. Because we perform 16 independent current injections, in total, $208$ $(= 16 \times 13)$ voltages are obtained and used to produce the CVS.

\subsection{CVS Extraction Using EIT and Influence of Motion}
A transconductance (column) vector $\boldsymbol g_t \in \mathbb{R}^{208}$ can be defined using the voltage data \eqref{voltagedata} as follows:
\begin{equation}\label{conductance}
	{\footnotesize \boldsymbol g_t = \begin{bmatrix} \dfrac{I}{\mathfrak R(V^{1,3}_t)}, \cdots, \dfrac{I}{\mathfrak R(V^{1,15}_t)}, \cdots, \dfrac{I}{\mathfrak R(V^{16,2}_t)},\cdots  \dfrac{I}{\mathfrak R(V^{16,14}_t)} \end{bmatrix}^T}
\end{equation}
where $T$ represents the vector transpose and $\mathfrak R$ is an operation for extracting the real part of a complex number. Here, $\boldsymbol g_t$ is updated every 10ms. 

A CVS, denoted by $\boldsymbol x_t \in \mathbb{R}$, is obtained by 
\begin{equation}
	\boldsymbol x_t = \boldsymbol w^T \dot{\boldsymbol g_t}
\end{equation}
where $\boldsymbol w \in \mathbb{R}^{208}$ is a weighting (so-called leadforming) vector and $\dot{\boldsymbol g_t}$ is time difference of $\boldsymbol g_t$ given by
\begin{equation}
	\dot{\boldsymbol g}_t = \boldsymbol g_t - \boldsymbol g_{t_0} \mbox{ for reference time } t_0 
\end{equation}
In the absence of motion, the transconductance $\dot{\boldsymbol g}_t$ can be expressed by
\begin{equation}\label{noMAcondi}
	\dot{\boldsymbol g}_t = \dot{\boldsymbol g}^{\mbox{\scriptsize air}}_t + \dot{\boldsymbol g}^{\mbox{\scriptsize blood}}_t
\end{equation}
where ${\boldsymbol g}^{\mbox{\scriptsize air}}_t$ and ${\boldsymbol g}^{\mbox{\scriptsize blood}}_t$ are transconductance vectors related to air and blood volume changes in the lungs and heart, respectively. The weighting vector $\boldsymbol w$ is designed to provide
\begin{equation}\label{LFcondi}
	\boldsymbol w^T \dot{\boldsymbol g}_t =  \boldsymbol w^T(\dot{\boldsymbol g}^{\mbox{\scriptsize air}}_t + \dot{\boldsymbol g}^{\mbox{\scriptsize blood}}_t)= \boldsymbol w^T \dot{\boldsymbol g}^{\mbox{\scriptsize blood}}_t
\end{equation}
See Figure \ref{EItsystem}. Kindly refer to \cite{Lee2021} for details on determining $\boldsymbol w$. Even though the cardiogenic signal ${\boldsymbol g}^{\mbox{\scriptsize blood}}_t$ is weak, it can be accurately decomposed from the data $\boldsymbol g_t$.

In light of the previous analysis in \cite{Lee2017}, the following explains why the quality of the CVS is degraded by motion, as shown in the middle part of Figure \ref{CVSmotionartifacts}. In the presence of motion, the transconductance $\dot{\boldsymbol g}_t$ can be approximated by
\begin{equation}\label{motionartifacts}
	\dot{\boldsymbol g}_t \approx \dot{\boldsymbol g}_t^{\mbox{\scriptsize normal}} + \dot{\boldsymbol g}^{\mbox{\scriptsize motion}}_t
\end{equation} 
where $\dot{\boldsymbol g}^{\mbox{\scriptsize normal}}_t = \dot{\boldsymbol g}^{\mbox{\scriptsize air}}_t + \dot{\boldsymbol g}^{\mbox{\scriptsize blood}}_t$ and $\dot{\boldsymbol g}^{\mbox{\scriptsize motion}}_t$ is the motion-induced effect. Appendix \ref{appendixA} presents details of \eqref{motionartifacts}. Determining the vector $\boldsymbol w$ itself can be considerably affected by motion artifacts \cite{Lee2021}. Moreover, even if $\boldsymbol w$ satisfies \eqref{LFcondi}, we have
\begin{equation}
	\boldsymbol x_t = \boldsymbol w^T \dot{\boldsymbol g}_t \approx \boldsymbol x_t^{\mbox{\scriptsize normal}} + \boldsymbol x_t^{\mbox{\scriptsize motion}}
\end{equation}
where $\boldsymbol x_t^{\mbox{\scriptsize normal}} = \boldsymbol w^T \dot{\boldsymbol g}_t^{\mbox{\scriptsize blood}}$ and $\boldsymbol x_t^{\mbox{\scriptsize motion}} = \boldsymbol w^T \dot{\boldsymbol g}_t^{\mbox{\scriptsize motion}}$. The last term $\boldsymbol x_t^{\mbox{\scriptsize motion}}$ describes motion artifacts in the CVS.

\begin{figure}[h]
	\centering
	\includegraphics[width=1\textwidth]{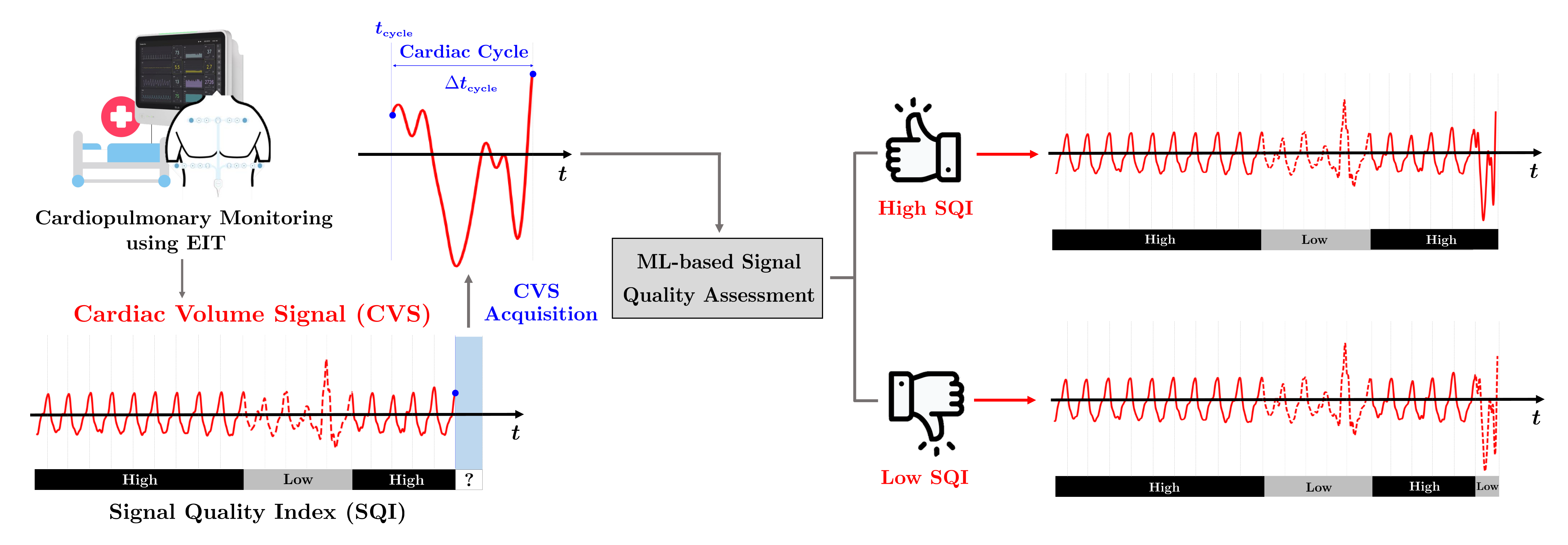}
	\caption{Schematic description of machine learning-based signal quality assessment for cardiac volume monitoring in electrical impedance tomography.}
	\label{mainfig}
\end{figure}

\subsection{CVS Quality Assessment and Data Preprocessing} \label{DataPre}
This study aims to assess the CVS ($\boldsymbol x_t$) for detecting motion-induced signal quality degradation. See Figure \ref{mainfig}. This can be accomplished by developing an SQI map $\boldsymbol f : \boldsymbol x_t \mapsto \boldsymbol y_t$ such that
\begin{equation} \label{goal1}
	\boldsymbol f(\boldsymbol x_t) = \boldsymbol y_t  = \left\{ \begin{array}{cl} 1 & \mbox{if } ~ \boldsymbol x_t^{\mbox{\scriptsize motion}} \approx 0 \\ 0 & \mbox{if } ~ \boldsymbol x_t^{\mbox{\scriptsize motion}} \not\approx 0 \\ \end{array}\right.
\end{equation} 
However, it is arduous to achieve \eqref{goal1}, where the assessment is conducted on an individual CVS at every sampling time. Instead, we take advantage of the periodicity and regularity of cardiac volume changes according to the heartbeat. The time interval of each cardiac cycle is identified using a synchronized ECG system.

Our quality assessment is conducted on every cardiac cycle of CVS, where a cardiac cycle is defined by the time interval consisting of two consecutive ECG R-wave peaks as the end points. For a given time $t_{\mbox{\scriptsize cyc}}$, let the interval $[t_{\mbox{\scriptsize cyc}},t_{\mbox{\scriptsize cyc}}+\Delta t_{\mbox{\scriptsize cyc}}]$ be the corresponding cardiac cycle, where $\Delta t_{\mbox{\scriptsize cyc}}$ is assumed to be $\Delta t_{\mbox{\scriptsize cycle}} = 10\mbox{ms} \times (\boldsymbol v-1)$ for some $\boldsymbol v \in \mathbb{N} \backslash \{1\}$. Here, $\mathbb{N}$ denotes the set of positive integers. A vector gathering all CVSs during the cycle, denoted by $\boldsymbol X_{t_{\mbox{\scriptsize cyc}}} \in \mathbb{R}^{\boldsymbol v}$, is defined as
\begin{equation}\label{CVSvec}
	\boldsymbol X_{t_{\mbox{\scriptsize cyc}}} = \begin{bmatrix} \boldsymbol x_{t_{\mbox{\scriptsize cyc}}}, \boldsymbol x_{t_{\mbox{\scriptsize cyc}}+10\mbox{\scriptsize ms}}, \cdots, \boldsymbol x_{t_{\mbox{\scriptsize cyc}}+10\mbox{\scriptsize ms} \times (\boldsymbol v-1)} \end{bmatrix} ^T
\end{equation}
The map $f$ in \eqref{goal1} can be modified into
\begin{equation} \label{goal2}
	\boldsymbol f(\boldsymbol X_{t_{\mbox{\scriptsize cyc}}}) = \boldsymbol y_{t} = \left\{ \begin{array}{cl} 1 & \mbox{for normal } \boldsymbol X_{t_{\mbox{\scriptsize cyc}}} \\ 0 & \mbox{for motion-influenced }\boldsymbol X_{t_{\mbox{\scriptsize cyc}}} \\\end{array}\right.
\end{equation}

\begin{figure}[h]
	\centering
	\includegraphics[width=1\textwidth]{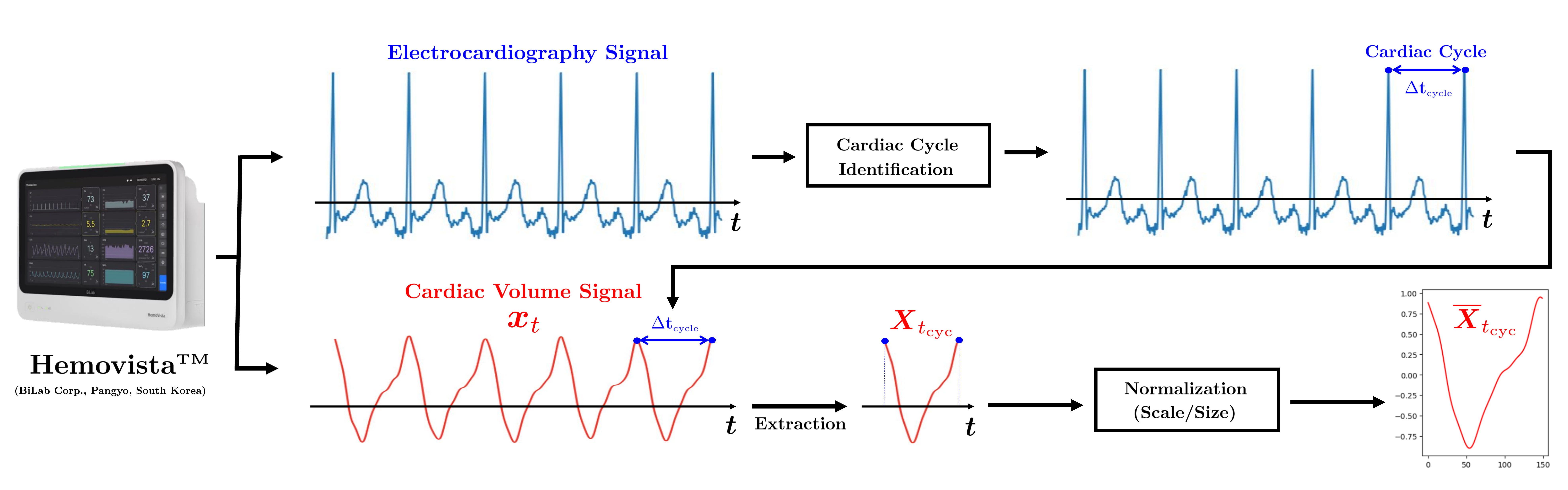}
	\caption{From the monitoring system, electrocadiography and cardiac volume signals are obtained. By identifying a cardiac cycle through electrocadiography data (R-wave peak detection), we extract cardiac volume signals at the corresponding cycle and then lastly apply normalization in terms of scale and size.}
	\label{CVSNormal}
\end{figure}

To find $\boldsymbol f$ in \eqref{goal2}, we leverage ML, which can learn the domain knowledge of normal and motion-influenced CVSs from a training dataset of $N$ data pairs $\{\boldsymbol X^{(i)}, \boldsymbol y^{(i)}\}_{i=1}^{N}$. Prior to ML applications, the following issues need to be addressed in the CVS data. First, CVSs have significant inter-subject and intra-subject variability. This is because cardiac volume varies depending on various factors, including sex, age, condition, time, and body temperature. Therefore, scale normalization is required to enhance the stability and performance of ML while mitigating the high learning complexity associated with scale-invariant feature extraction \cite{Goodfellow2016,Xu2014}. Second, the dimensions of the input CVS data in \eqref{goal2} do not match each other (i.e., $\boldsymbol v$ is not constant) owing to heart rate variability \cite{Conny1993}. Because most existing ML methods are based on an input with consistent dimensions, size normalization is required. Figure \ref{CVSNormal} schematically illustrates the overall process.

\subsubsection{Scale normalization}
A simple method of normalizing the scale is to rescale the CVS data for individual cardiac cycles. Specifically, for a given CVS vector $\boldsymbol X_{t_{\mbox{\scriptsize cyc}}} \in \mathbb{R}^{\boldsymbol v}$, the scaling factor $\mathcal S$ is obtained using
\begin{equation}\label{simplerescale}
	\mathcal S = \underset{i \in \mathcal V}{\mbox{max}} ~ |\boldsymbol x_{t_{\mbox{\tiny cyc}}+10\times i\mbox{\tiny (ms)}}|
\end{equation}
where the index set $\mathcal V$ is given by $\mathcal V=\{0,1,\cdots, \boldsymbol v - 1\}$. Normalized CVS data, denoted by $\underline{\boldsymbol X}_{t_{\mbox{\tiny cyc}}}$, are obtained by
\begin{equation}\label{ScaleNormalization}
	\underline{\boldsymbol X}_{t_{\mbox{\tiny cyc}}} = \dfrac{\boldsymbol X_{t_{\mbox{\tiny cyc}}}}{\mathcal S}
\end{equation}
However, this scaling may not be appropriate to our application for the following reason. Abnormalities in CVS data include sudden increases or decreases in signal amplitude as well as irregular deformations of the shape profile. The normalization in \eqref{ScaleNormalization} can contribute to ignoring rapid amplitude changes.

This study uses the following subject-specific scale normalization strategy. When the EIT device is used to monitor a certain subject, it is supposed that during the initial 20s calibration process, the device measures the normal CVS data available for scale normalization. Let $\mathcal X^{\mbox{\scriptsize subject}}$ be a set of corresponding CVSs given by
\begin{equation}
	\mathcal X^{\mbox{\scriptsize subject}}=\{ \boldsymbol x_{10\times i\mbox{\tiny (ms)}} :  i = -1999, -1998, \cdots, -1 \}
\end{equation}
Using the set $\mathcal X^{\mbox{\scriptsize subject}}$, a subject-specific scaling factor $\mathcal S^{\mbox{\scriptsize subject}}$ is obtained by
\begin{equation}
	\mathcal S^{\mbox{\scriptsize subject}} = \underset{\boldsymbol x \in \mathcal X^{\mbox{\scriptsize subject}}}{\mbox{max}}|\boldsymbol x|
\end{equation}
This scale factor $\mathcal S^{\mbox{\scriptsize subject}}$ is used for the normalization in \eqref{ScaleNormalization} instead of the naive factor $\mathcal S$ in \eqref{simplerescale}.

\subsubsection{Size normalization}
To make the dimensions of the CVS data consistent, a CVS vector $\underline{\boldsymbol X}_{t_{\mbox{\scriptsize cyc}}}$ is embedded into $\mathbb{R}^{\nu}$ for a fixed constant $\boldsymbol \nu$. In the empirical experiment, the embedding space dimension was to be larger than any dimension of the CVS data in our dataset ($\boldsymbol \nu = 150$). 

Two normalization methods are considered. The first approach is to resample $\boldsymbol \nu$ points using linear interpolation with $\boldsymbol v$ data points in $\underline{\boldsymbol X}_t$. For the stationary interval $[0,1]$, the following linear interpolation function $\mathcal L$ is constructed: 
\begin{equation}
	\mathcal L(\dfrac{i}{\boldsymbol v-1}) = \underline{\boldsymbol x}_{t_{\mbox{\tiny cyc}}+10\mbox{\tiny (ms)} \times (i-1)}  \mbox{ for } i = 0,\cdots,\boldsymbol v-1
\end{equation}
Subsequently, we obtain the normalized vector $\overline{\boldsymbol X}_{t_{\mbox{\tiny cyc}}} \in \mathbb{R}^{\boldsymbol \nu}$ using
\begin{equation}
	\overline{\boldsymbol X}_{t_{\mbox{\tiny cyc}}} = \left[ ~ \mathcal L(0), \mathcal L(\dfrac{1}{\boldsymbol \nu - 1}), \mathcal L(\dfrac{2}{\boldsymbol \nu - 1}), \cdots, \mathcal L(1) ~ \right]^T
\end{equation}
This method normalizes the signal profile of CVS data into the desired length ($\boldsymbol \nu$) with no significant loss, but loses sampling time information. Second, the last value in $\underline{\boldsymbol X}_{t_{\mbox{\scriptsize cyc}}}$ (i.e., $\boldsymbol x_{t_{\mbox{\scriptsize cyc}}+10\mbox{\scriptsize (ms)} \times (\boldsymbol v-1)}$) is padded up to the desired length. This constant padding provides a vector $\overline{\boldsymbol X}_{t_{\mbox{\scriptsize cyc}}} \in \mathbb{R}^{\boldsymbol \nu}$, expressed by
\begin{align}
	\overline{\boldsymbol X}_{t_{\mbox{\tiny cyc}}} = &[ ~ \underline{\boldsymbol x}_{t_{\mbox{\tiny cyc}}}, \cdots, \underline{\boldsymbol x}_{t_{\mbox{\tiny cyc}}+10\mbox{\tiny (ms)} \times (\boldsymbol v-2)}, \underline{\boldsymbol x}_{t_{\mbox{\tiny cyc}}+10\mbox{\tiny (ms)} \times (\boldsymbol v-1)},  \\ & ~~ \underline{\boldsymbol x}_{t_{\mbox{\tiny cyc}}+10\mbox{\tiny (ms)} \times (\boldsymbol v-1)}, \cdots, \underline{\boldsymbol x}_{t_{\mbox{\tiny cyc}}+10\mbox{\tiny (ms)} \times (\boldsymbol v-1)} ~ ]^T\label{PADD}
\end{align} 
where the part \eqref{PADD} corresponds to the padding. In contrast to the first method, this normalization can preserve time information regarding sampling frequency, whereas the core profile of the CVS is supported at different time intervals. 

\subsection{Machine Learning Application}
At this point, we are ready to apply ML for determining the SQI function \eqref{goal2}. Collected from various subjects and cardiac cycles, the following dataset is used:
\begin{equation}\label{dataset}
	\{ \overline{\boldsymbol X}^{(i)}, \boldsymbol y^{(i)}\}_{i=1}^{N}
\end{equation}
where $\boldsymbol y^{(i)}$ is the SQI label corresponding to $\overline{\boldsymbol X}^{(i)}$. We note that $\overline{\boldsymbol X}$ is the CVS data for a cardiac cycle of some subjects and is normalized for both scale and size. In practice, the available training dataset \eqref{dataset} was highly imbalanced, where there were relatively few negative samples (motion-influenced CVSs).   

\begin{figure}[h]
	\centering
	\includegraphics[width=0.975\textwidth]{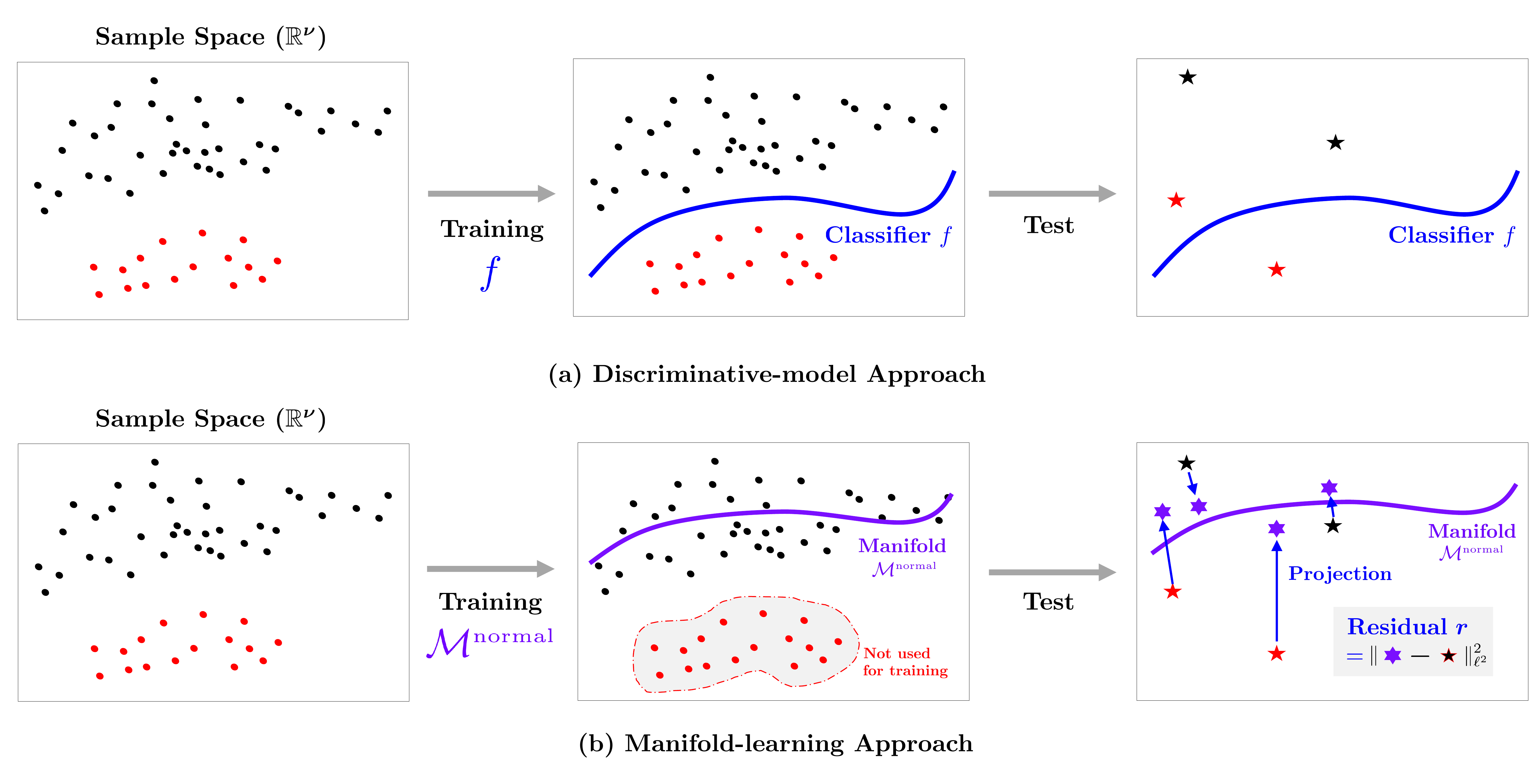}
	\caption{(a) Discriminative-model approach learns a signal quality indexing map $f$ by using CVS and label data. (b) Manifold-learning approach first learns common features of normal CVS data by finding a low dimensional manifold $\mathcal M^{\mbox{\scriptsize normal}}$. Signal quality assessment is based on computing the residual between original CVS data and projected one onto or near the learned manifold.}
	\label{MLFig}
\end{figure}

\subsubsection{Discriminative-model approach}
The discriminative-model approach trains the SQI map $\boldsymbol f : \overline{\boldsymbol X} \mapsto \boldsymbol y$ in the following sense:
\begin{equation}
	\boldsymbol f = \underset{\boldsymbol f \in \mathbb{F}}{\mbox{argmin}} ~ \dfrac{1}{N}\sum_{i=1}^{N}\mbox{dist}(\boldsymbol f(\overline{\boldsymbol X}^{(i)}),\boldsymbol y^{(i)})
\end{equation}
where $\mathbb{F}$ is a set of learnable functions for a given ML model and $\mbox{dist}$ is a metric that measures the difference between the ML output $f(\overline{\boldsymbol X})$ and label $\boldsymbol y$.  See Figure \ref{MLFig} (a). In our application with high class-imbalance, the following weighted cross-entropy can be used:
\begin{equation}
	\mbox{dist}(\boldsymbol f(\overline{\boldsymbol X}),\boldsymbol y) = - \boldsymbol \zeta_{\mbox{\scriptsize pos}}\boldsymbol y\mbox{log}(\boldsymbol f(\overline{\boldsymbol X}))- \boldsymbol \zeta_{\mbox{\scriptsize neg}}(1-\boldsymbol y) \mbox{log}(1-\boldsymbol f(\overline{\boldsymbol X}))  
\end{equation}
where $\boldsymbol \zeta_{\mbox{\scriptsize pos}}$ and $\boldsymbol \zeta_{\mbox{\scriptsize neg}}$ are the relative ratios of the positive and negative samples, respectively. Various classification models can be used, such as the logistic regression model (LR) \cite{Cramer2002}, multi-layer perceptron (MLP) \cite{Hinton2007}, and convolutional neural networks (CNN) \cite{Simonyan2014}. Detailed models used in this study are explained in Appendix \ref{appendixB1}.

The discriminative model approach is a powerful method to guarantee high performance in a fixed dataset. However, it might suffer from providing stable SQI results in clinical practice because of highly variable negative samples. This is because these methods take advantage of learned information using only a few negative samples \cite{Buda2018,Cao2019,He2009,Van2017}. To achieve stable prediction, the manifold-learning approach can be alternatively used \cite{Chapelle2009,Van2020,Zhu2009}. 

\subsubsection{Manifold-learning approach} \label{sec:semi-sup}
The manifold-learning approach learns common features from positive samples (i.e., normal CVS) and uses them to develop an SQI map. The remaining negative samples are utilized as auxiliary means for selecting a hyperparameter. Figure \ref{MLFig} (b) shows a schematic description of this process. 

A set of positive samples is denoted by $\{ \overline{\boldsymbol X}_{\mbox{\scriptsize pos}}^{(i)}\}_{i=1}^{N_{\mbox{\scriptsize pos}}}$, where $N_{\mbox{\scriptsize pos}}$ denotes the number of positive samples. In the first step, we learn a low-dimensional representation of $\overline{\boldsymbol X}_{\mbox{\scriptsize pos}}$ by training an encoder $\mathcal E : \overline{\boldsymbol X}_{\mbox{\scriptsize pos}} \mapsto \boldsymbol z$ and decoder $\mathcal D : \boldsymbol z \mapsto \overline{\boldsymbol X}_{\mbox{\scriptsize pos}}$ in the following sense \cite{Jolliffe2016,Hinton2006}:
\begin{equation}\label{AELoss}
	(\mathcal D, \mathcal E) = \underset{(\mathcal D, \mathcal E)}{\mbox{argmin}} ~ \dfrac{1}{N_{\mbox{\scriptsize pos}}}\sum_{i=1}^{N_{\mbox{\scriptsize pos}}}\| \mathcal D \circ \mathcal E (\overline{\boldsymbol X}_{\mbox{\scriptsize pos}}^{(i)}) - \overline{\boldsymbol X}_{\mbox{\scriptsize pos}}^{(i)} \|_{2}^2
\end{equation}
where $\boldsymbol z$ is a low dimensional latent vector and $\|\cdot\|_2$ is the standard Euclidean norm. The architectures $\mathcal D$ and $\mathcal E$ can be used in PCA \cite{Jolliffe2016}, VAE \cite{Kingma2013}, and $\beta$-VAE \cite{Higgins2016}. See more details in Appendix \ref{appendixB2}. 

Borrowing the idea from \cite{An2015}, an SQI map $f$ is constructed as follows: For a given CVS data $\overline{\boldsymbol X}$ in any class, a residual $\boldsymbol r$ is computed by
\begin{equation}\label{residual}
	\boldsymbol r = \| \overline{\boldsymbol X} - \mathcal D \circ \mathcal E(\overline{\boldsymbol X})\|_{2}
\end{equation}
The decoder $\mathcal D$ is trained to generate normal CVS-like output. In other words, operation $\mathcal D \circ \mathcal E$ transforms $\overline{\boldsymbol X}$ to lie in or near the learned manifold using normal CVS data \cite{Seo2013,Yun2022}. Therefore, the residual $\boldsymbol r$ can be viewed as an anomaly score, where $\boldsymbol r$ is small if $\overline{\boldsymbol X}$ is normal CVS data, and large if $\overline{\boldsymbol X}$ is motion-influenced CVS data. For some non-negative constant $d$, an SQI map $\boldsymbol f$ can be constructed using
\begin{equation}\label{selfsupclassi}
	\boldsymbol f(\overline{\boldsymbol X}_t) = \left \{ \begin{array}{cl} 1 & \mbox{if } \boldsymbol r \leq  \boldsymbol d \\ 0 & \mbox{if } \boldsymbol r > \boldsymbol d \end{array} \right.
\end{equation}

The remainder of this subsection explains how the thresholding value $\boldsymbol d$  is determined by utilizing negative samples as well as positive. By varying $\boldsymbol d$ from 0 to $\infty$, a receiver operating characteristic (ROC) curve is calculated, where a point in the ROC curve is obtained using a fixed $\boldsymbol d$. We choose $\boldsymbol d$ such that maximizing Youden's $\boldsymbol J$ statistics, which is known as an unbiased metric in the class imbalance case \cite{Ruopp2008}. The value $\boldsymbol J$ is given by
\begin{equation} \label{Jstatictics}
	\boldsymbol J^{\boldsymbol d}  = \mbox{Sensitivity}^{\boldsymbol d} + \mbox{Specificity}^{\boldsymbol d} - 1 
\end{equation}
where
\begin{equation}
	\mbox{Sensitivity}^{\boldsymbol d} = \dfrac{N_{\mbox{\scriptsize TP}}^{\boldsymbol d}}{N_{\mbox{\scriptsize TP}}^{\boldsymbol d}+N_{\mbox{\scriptsize FN}}^{\boldsymbol d}} \mbox{ and }\mbox{Specificity}^{\boldsymbol d} = \dfrac{N_{\mbox{\scriptsize TN}}^{\boldsymbol d}}{N_{\mbox{\scriptsize TN}}^{d}+N_{\mbox{\scriptsize FP}}^{\boldsymbol d}}
\end{equation}  
Here, $N_{\mbox{\scriptsize TP}}^{\boldsymbol d}$, $N_{\mbox{\scriptsize TN}}^{\boldsymbol d}$, $N_{\mbox{\scriptsize FP}}^{\boldsymbol d}$, and $N_{\mbox{\scriptsize FN}}^{\boldsymbol d}$ respectively represent the number of true positives, true negatives, false positives, and false negatives for predictions depending on a selected threshold value $\boldsymbol d$.

\section{Results}
\subsection{Data Acquisition and Experimental Setting}
Our dataset was obtained from healthy volunteers using an EIT-based hemodynamic monitoring device (HemoVista, BiLab, South Korea). Synchronized ECG data were obtained with EIT and used to identify the cardiac cycles. While lying in a hospital bed, each subject was requested to make intentional motions mimicking postural changes in the clinical ward. A total of 16140 CVS data were obtained regarding the cardiac cycle.

Manual labeling was individually performed by two- and ten- years bio-signal experts (Nam and Lee). Subsequently, they reviewed the results and made the final decision about CVS abnormality through an agreement between them. The final labels were annotated into three classes: normal, ambiguous, and motion-influenced. When classified as normal or abnormal by both experts with an agreement, CVS data were annotated as normal or motion-influenced classes. The ambiguous class stands for CVS data in which motion artifacts were included with high possibility, but the experts did not reach an explicit agreement about motion influence. The assigned label is $\boldsymbol y = 1$ for the normal class and $\boldsymbol y = 0$ for the other classes. As a result, 12928 (80.09$\%$), 1526 (9.45$\%$), and 1686 (10.45$\%$) samples were labeled as normal, ambiguous, and motion-influenced classes, respectively. 

For ML applications, a total of 16372 CVS data were divided into 13100 (80$\%$), 1520 (10$\%$), and 1520 (10$\%$), which were used for training, validation, and testing, respectively. The data split was performed such that CVS data obtained from a common subject did not exist between the three sets. For the training dataset, labels for the ambiguous class were reassigned to $\boldsymbol y = 0.25$. This was done to prevent the over-classification of ambiguous classes.

ML experiments were conducted in a computer system with GeForce RTX 3080 Ti, Intel® Core™ X-series Processors i9-10900X, and 128GB DDR4 RAM. Python with scikit-learn and Pytorch packages were used for the ML implementation. When training the ML models, the Adam optimizer was consistently employed, which is an effective adaptive stochastic gradient descent method \cite{Kingma2014}. Hyperparameters such as epoch and learning rate were heuristically chosen based on the validation results.

\subsection{Results of CVS Quality Assessment}
We compared the performance of the ML-based CVS quality assessment results by using six metrics: accuracy, positive and negative predictive values (PPV and NPV), sensitivity, specificity, and AUC. Accuracy, PPV, and NPV were defined by
\begin{equation}
	\mbox{Accuracy} = \dfrac{N_{\mbox{\scriptsize TP}} + N_{\mbox{\scriptsize TN}}}{N_{\mbox{\scriptsize TP}}+N_{\mbox{\scriptsize TN}}+N_{\mbox{\scriptsize FP}}+N_{\mbox{\scriptsize FN}}}, \mbox{PPV} = \dfrac{N_{\mbox{\scriptsize TP}}}{N_{\mbox{\scriptsize TP}}+N_{\mbox{\scriptsize FP}}}, \mbox{ and } \mbox{NPV} = \dfrac{N_{\mbox{\scriptsize TN}}}{N_{\mbox{\scriptsize TN}}+N_{\mbox{\scriptsize FN}}}
\end{equation}
and AUC was the area under the ROC curve. NPV, specificity, and AUC should be emphasized in our evaluation owing to the high-class imbalance (small negative samples).

\begin{table}[h]
	{\scriptsize
	\begin{center}
		\textbf{(a) SQI with scale and size normalization using linear interpolation.}
		\begin{tabular}{|p{1.5cm}|p{1.5cm}|p{1.5cm}|p{1.5cm}|p{1.5cm}|p{1.5cm}|p{1.5cm}|p{1.5cm}|} \hline
			\multicolumn{2}{|c|}{\textbf{Discriminative Model}} & \textbf{LR} & \textbf{MLP1} & \textbf{MLP2} & \textbf{VGG16-3} & \textbf{VGG16-4} & \textbf{VGG16-5} \\ \hline
			\multirow{6}{*}{\textbf{Test}} & \textbf{Accuracy} & 0.8665 & 0.9323 & 0.9348 & 0.9468 & 0.9468 & 0.9437 \\ 
			& \textbf{PPV} & 1.0000 & 0.9790 & 0.9747 & 0.9525 & 0.9605 & 0.9679 \\ 
			& \textbf{NPV} & 0.1097 & 0.7241 & 0.7445 & 0.9047 & 0.8591 & 0.8083 \\ 
			& \textbf{Sensitivity} & 0.8643 & 0.9404 & 0.9479  & 0.9866 & 0.9776 & 0.9657 \\ 
			& \textbf{Specificity} & 1.0000 & 0.8860 & 0.8607 & 0.7215 & 0.7721 & 0.8185 \\
			& \textbf{AUC} & 0.6615 & 0.9506 & 0.9558 & 0.9709 & 0.9645 & 0.9653 \\ \hline
			\multicolumn{2}{|c|}{\textbf{Manifold-learning Model}} & \textbf{PCA} & \textbf{VAE} & \textbf{$\boldsymbol \beta$-VAE} & \textbf{CVAE} & \textbf{$\boldsymbol \beta$-CVAE} & - \\ \hline
			\multirow{6}{*}{\textbf{Test}} & \textbf{Accuracy} & 0.8468 & 0.9066 & 0.9221 & 0.9292 & 0.9298  &\\ 
			& \textbf{PPV} & 0.9510 & 0.9687 & 0.9672 & 0.9688 & 0.9739 & \\ 
			& \textbf{NPV} & 0.4573  & 0.6181 & 0.6900 & 0.7100 & 0.7011 & \\ 
			& \textbf{Sensitivity} & 0.8675 & 0.9218 & 0.9439 & 0.9486 & 0.9441 & -\\ 
			& \textbf{Specificity} & 0.7142 & 0.8095 & 0.7952 & 0.8047 & 0.8380 & \\ 
			& \textbf{AUC} & 0.8735 & 0.9513 & 0.9489 & 0.9528 & 0.9603 & \\ \hline
		\end{tabular}
	\end{center}
	\begin{center}
		\textbf{(b) SQI with scale and size normalization using constant padding.}
		\begin{tabular}{|p{1.5cm}|p{1.5cm}|p{1.5cm}|p{1.5cm}|p{1.5cm}|p{1.5cm}|p{1.5cm}|p{1.5cm}|} \hline
			\multicolumn{2}{|c|}{\textbf{Discriminative Model}} & \textbf{LR} & \textbf{MLP1} & \textbf{MLP2} & \textbf{VGG16-3} & \textbf{VGG16-4} & \textbf{VGG16-5} \\ \hline
			\multirow{6}{*}{\textbf{Test}} & \textbf{Accuracy} & 0.8664 & 0.9487 & 0.9518 & 0.9455 & 0.9487 & 0.9500 \\ 
			& \textbf{PPV} & 1.0000 & 0.9745 & 0.9767 & 0.9533 & 0.9655 & 0.9731 \\ 
			& \textbf{NPV} & 0.0826 & 0.7851 & 0.8065 & 0.8870 & 0.8433 & 0.8185 \\ 
			& \textbf{Sensitivity} & 0.8648 & 0.9651 & 0.9666 & 0.9844 & 0.9748 & 0.9681 \\ 
			& \textbf{Specificity} & 1.0000 & 0.8521 & 0.8652 & 0.7173 & 0.7956 & 0.8434 \\
			& \textbf{AUC} & 0.6628 & 0.9725 & 0.9669 & 0.9782 & 0.9683 & 0.9757 \\ \hline
			\multicolumn{2}{|c|}{\textbf{Manifold-learning Model}} & \textbf{PCA} & \textbf{VAE} & \textbf{$\boldsymbol \beta$-VAE} & \textbf{CVAE} & \textbf{$\boldsymbol \beta$-CVAE} & - \\ \hline
			\multirow{6}{*}{\textbf{Test}} & \textbf{Accuracy} & 0.8809 & 0.8918 & 0.9214 & 0.9015 & 0.8861 &\\ 
			& \textbf{PPV} & 0.9590 & 0.9660 & 0.9679 & 0.9731 & 0.9636 & \\ 
			& \textbf{NPV} & 0.5333 & 0.5629 & 0.6694 & 0.5882 & 0.5467 & \\ 
			& \textbf{Sensitivity} & 0.9014 & 0.9074 & 0.9407 & 0.9118 & 0.9029 & -\\ 
			& \textbf{Specificity} & 0.7450 & 0.7892 & 0.7941 & 0.8333 & 0.7745 & \\ 
			& \textbf{AUC} & 0.9150 & 0.9206 & 0.9412 & 0.9170 & 0.9041 & \\ \hline
		\end{tabular}
	\end{center}}
	\caption{Machine learning-based CVS quality assessment results}	\label{result}
\end{table}

\subsubsection{Discriminative Models} 
The first and second rows of Tables \ref{result} (a) and (b) show the quantitative evaluations of CVS quality assessment using various discriminative models: LR, MLPs, and CNNs. The results in Tables \ref{result} (a) and (b) differ in size normalization: (a) linear interpolation and (b) constant padding. 

MLPs and CNNs performed better than LR, which provided miserable NPV and AUC. MLPs and CNNs outperformed each other in specificity and NVP respectively, while achieving comparable levels for the other metrics. There was no significant performance gap depending on the size normalization.

One interesting observation was as follows: In our experiments, there seems to be a compensation between specificity and NPV, depending on the emphasis on locality and globality. Enriching global information on CVS data positively affected specificity; in contrast, local information helped improve NPV. As the receptive field size in VGG16 increased (see Appendix \ref{appendixB1}), specificity tended to increase and NPV decrease. In MLP, which is more flexible for catching global information than CNNs, specificity was highest, and NPV lowest. In other words, the local information of CVS data is likely to play a crucial role in reducing false negatives rather than false positives. From a practical point of view, reducing false negatives is more desirable; therefore, using VGG16-3 or VGG16-4, which have the powerful ability to take advantage of locality, can be an excellent option.

\subsubsection{Manifold-learning Models} 
\begin{figure}[t]
	\centering
	\includegraphics[width=1\textwidth]{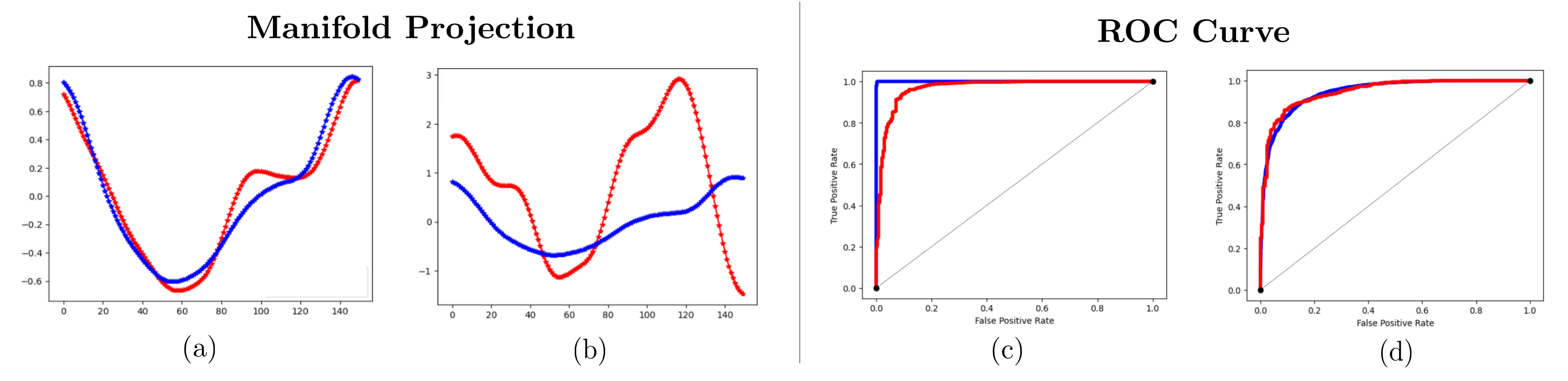}
	\caption{Test samples and VAE-based projection results for (a) normal and (b) motion-influenced CVS data, where the red line is original CVS data and the blue line is the correspondent CVS data projected by VAE. By the way, ROC curves for (c) VGG 16-4 and (d) $\beta$-VAE are provided, where the blue and red lines correspond to the curves calculated using training and test sets, respectively.}
	\label{Result2}
\end{figure}
Positive samples in the validation set were used for hyperparameter selection in training the encoder and decoder. A threshold value was determined by using data from all the training and validation sets. 

Figure \ref{Result2} shows manifold projection results of test samples in normal and motion-influenced classes. An input CVS is projected onto or near a manifold learned by positive samples. As desired, the residual \eqref{residual} tends to be small for normal samples and high for motion-influenced samples.

The third and fourth rows of Tables \ref{result} (a) and (b) show the final assessment results using manifold-learning models. The performance was comparable to that of discriminative models. We note that the manifold-learning models never learned negative samples for classifier development. As shown in Figure \ref{Result2} (d), the manifold-learning model's performance gap between training and test sets was very small.

There was a slight difference in performance for the manifold-learning models depending on the size normalization. Linear interpolation promised a slightly better assessment of accuracy, NPV, and AUC than the other. For the case of constant padding, because core profiles of CVS data are supported at different intervals, the learning complexity can be increased, which is associated with invariant feature extraction to the intervals. This may cause a slight drop in performance.

In our dataset, both discriminative and manifold learning models provided accurate detection of motion-influenced CVS. The discriminative model yielded a more powerful SQI performance; in contrast, the manifold-learning model provided stable outcomes between the training and test sets. Regarding practical applications, the choice of two models relies on what should be emphasized in the monitoring system in terms of performance and stability. Their ensemble is also worth considering.

\subsubsection{Impact of Scale Normalization}
\begin{table}[h]
	\centering
	{\footnotesize 
		\begin{tabular}{|p{2cm}|p{1.75cm}p{1.4cm}|p{1.75cm}p{1.4cm}|} \hline
			& \multicolumn{2}{l|}{\textbf{With Scaling}} & \multicolumn{2}{l|}{\textbf{Without Scaling}} \\ \hline
			\textbf{Model} & \textbf{VGG16-3} & \textbf{VAE} & \textbf{VGG16-3} & \textbf{VAE} \\ \hline
			\textbf{Accuracy} & 0.9468 & 0.9066 & 0.7862 & 0.7509 \\
			\textbf{PPV} & 0.9525 & 0.9687 & 0.9763 & 0.9668 \\ 
			\textbf{NPV} & 0.9047 & 0.6181 & 0.4038 & 0.3327 \\
			\textbf{Sensitivity} & 0.9886 & 0.9218 & 0.7671 & 0.7373 \\
			\textbf{Specificity} & 0.7215 & 0.8095 & 0.8945 & 0.8380 \\
			\textbf{AUC} & 0.9709 & 0.9513 & 0.9067 & 0.8906 \\ \hline
	\end{tabular}}
	\caption{Results of machine learning-based CVS quality assessment with and without scale normalization.}
	\label{ScaleNormal}
\end{table}
 
Table \ref{ScaleNormal} shows the worst case when scale normalization was not applied. In CNNs, network training was very unstable, and assessment performance was considerably degraded, especially regarding accuracy, NPV, sensitivity, and AUC. In VAEs, large-scale variability of CVS data highly affected the loss of accuracy in manifold projection; therefore, the performance significantly deteriorated in terms of accuracy, NPV, sensitivity, and AUC. This verifies the impact of scale normalization.

\subsubsection{Inference Time} 
\begin{table}[h]
	\centering
	{\footnotesize 
		\begin{tabular}{|p{1.75cm}|p{1.75cm}|p{1.75cm}|p{1.75cm}|p{1.75cm}|p{1.75cm}|} \hline
			\textbf{Model} & \textbf{LR} & \textbf{MLP1} & \textbf{MLP2} & \textbf{VGG16-3} & \textbf{VGG16-4}  \\ \hline
			\textbf{Time} & 0.633$\mu$s & 1.265$\mu$s & 0.700$\mu$s & 1.897$\mu$s & 3.162$\mu$s \\ \hline
			\textbf{Model} & \textbf{VGG16-5} & \textbf{PCA} & \textbf{VAE} &\textbf{CVAE} & - \\ \hline
			\textbf{Time} & 3.562$\mu$s & 48.412$\mu$s & 3.703$\mu$s & 15.192$\mu$s & -  \\ \hline
	\end{tabular}}
	\caption{Test inference time of machine learning-based CVS quality assessment methods.}
	\label{InferenceTime}
\end{table}

In real-time monitoring, assessment should be performed quickly. The input for the proposed method was updated for every heartbeat in the EIT system. Assuming a subject with a constant 80bpm, the CVS input is updated every 0.75s. Roughly, the assessment should be faster than approximately $10^{-2}$s. Table \ref{InferenceTime} shows the inference time for the test data, calculated by taking the average over the entire test data. The ML models provided a test outcome with inference times between 100$\mu$s ($10^{-4}$s) and 0.1$\mu$s ($10^{-7}$s). This confirms that the proposed method meets the speed requirements for real-time monitoring.

\section{Conclusion and Discussion}
We developed a novel automated SQI method using two machine learning techniques, the discriminative model and manifold learning, to detect abnormal CVS caused by motion-induced artifacts. We discussed how body movement influences the transconductance data and how the resulting CVS is degraded by movement. Numerous experiments support the idea that the proposed method can successfuly filter motion-induced unrealistic variations in CVS data.

To the best of our knowledge, this is the first attempt to assess CVS quality to enhance the clinical capability of an EIT-based cardiopulmonary monitoring system. From a practical point of view, the proposed method can alert clinicians about CVS corruption to minimize misinformation about patient safety and facilitate adequate management of patients and medical resources. The proposed method can be combined with a software system for existing EIT devices.

The use of only healthy subject data in the training process did not fully consider possible influence of the subject's illness on CVS. SQI performance might be degraded in patients with illnesses such as arrhythmias, in which irregular deformation may occur in CVS due to premature ventricular contraction and lead to be classified as low signal quality. However, when ill patient data are available and appended in the training process, a slightly modified SQI can detect the illness and motion by adding another label class. Meanwhile, arrhythmia can be easily detected using ECG signals.

A further collection of CVS data could be a strategy for enhancing model generalization or stability toward being equipped with an actual monitoring system. In discriminative models, even with additional data collection, generalization or stability might not be meaningfully improved because the class imbalance problem remains or increases. In contrast, the manifold-learning models can accurately infer common features (i.e., data manifolds) as the total number of normal CVS data grows regardless of class imbalance. In addition, it can be extended into a semi-supervised or unsupervised learning framework \cite{An2015,Schlegl2017}, which reduces the requirement for labeled datasets. Thus, manifold-learning models might be favorable.

\section*{Data Availability}
The data that support the findings of this study are available from the corresponding author, K. Lee, upon reasonable request.

\section*{Acknowledgements}
This work was supported by the Ministry of Trade, Industry and Energy (MOTIE) in Korea through the Industrial Strategic Technology Development Program under Grant 20006024. Hyun was supported by Samsung Science \& Technology Foundation (No. SRFC-IT1902-09). We are deeply grateful to BiLab (Pangyo, South Korea) for their help and collaboration.

\section*{Conflict of Interest}
The authors have no conflicts to disclose. \\

\appendix
\section{Motion-induced Effect on Trans-conductance} \label{appendixA}
In the 16 channel EIT system, the voltage data $\{V^{j,k}_t\}_{j,k}$ in \eqref{voltagedata} are governed by the following complete electrode model \cite{Seo2013}: At time $t$, the electric potential distribution ($u_t^{j}$) and electric potential on an electrode ($U_t^{j,k}$) satisfy
\begin{equation}
	\left \{ \begin{array}{rl} \displaystyle 
		\nabla \cdot( \gamma_t \nabla u^j_t ) &= 0 ~~~~  \mbox{ in }\Omega \subset \mathbb{R}^{3} \\
		\gamma_t \nabla u_t^j \cdot n & = 0  ~~~~ \mbox{ on } \partial \Omega \backslash \bigcup_i^{16} \mathcal E^k \\ \displaystyle 
		\int_{\mathcal E^k} \gamma_t \nabla u_t^j \cdot n & = 0 ~~~~ \mbox{ for } k \in \mathcal I \backslash \{j,j+1\} \\ \displaystyle 
		u_t^j + z_k(\gamma_t \nabla u_t^j\cdot n) &= U_t^{j,k}  \mbox{ on } \mathcal E^k \mbox{ for } k \in \mathcal I \\ \displaystyle 
		\int_{\mathcal E^j} \gamma_t \nabla u_t^j \cdot n ds & = -\displaystyle \int_{\mathcal E^{j+1}} \gamma_t \nabla u_t^j \cdot n ds = I 
	\end{array} \right.
\end{equation}
where $\gamma_t$ is a conductivity distribution in a human chest $\Omega$ at $t$, $n$ is an unit normal vector outward $\partial \Omega$, $ds$ is a surface element, and $z_k$ is a skin-electrode contact impedance on $\mathcal E^k$. The amount of electric current $I$, which is injected to the domain $\Omega$, can be scaled and, thus, assumed to be $I=1$.

In the case that the human chest $\Omega$ is time-varying owing to motions, Reynolds transport theorem yields the following approximation \cite{Lee2021}:  
\begin{equation}\label{VoltageMotion}
	\dot{V}^{j,k}_t \approx \dot{V}^{j,k,\mbox{\scriptsize normal}}_{t} + \dot{V}^{j,k,\mbox{\scriptsize motion}}_{t}
\end{equation}
where 
\begin{align}
	\dot{V}^{j,k,\mbox{\scriptsize normal}}_{t} & = - \int_{\Omega} \dot{\gamma}_t(\boldsymbol r) \nabla u^j_t(\boldsymbol r) \cdot \nabla u_t^k(\boldsymbol r) d\boldsymbol r \\
	\dot{V}^{j,k,\mbox{\scriptsize motion}}_{t} & = - \int_{\partial \Omega} \boldsymbol v_n(\boldsymbol r,t) \gamma_t(\boldsymbol r) \nabla u^j_t(\boldsymbol r)\cdot \nabla u^{k}_t(\boldsymbol r) ds \label{VGMotion}
\end{align}
Here, $\boldsymbol v_n$ is an outward-normal directional velocity of $\partial \Omega$ and $\boldsymbol r \in \Omega$ is a position vector in $\Omega$. The term $\dot{V}^{j,k,\mbox{\scriptsize normal}}_{t}$ and $\dot{V}^{j,k,\mbox{\scriptsize motion}}_{t}$ can be viewed as voltage data acquirable in normal EIT measurement and motion-induced inference, respectively.

A similar relation to \eqref{VoltageMotion} for trans-conductance can be derived as follows: Let us define a trans-conductance-related value $g_t^{j,k}$ by 
\begin{equation}
	g_t^{j,k}=\dfrac{I}{\mathfrak R(V_t^{j,k})}
\end{equation}
By differentiating $g_t^{j,k}$ with respect to $t$, we obtain
\begin{equation}\label{transconductance_derivative}
	{\scriptsize \dot{g}_t^{j,k}= \dfrac{ -I \mathfrak R(\dot{V}_t^{j,k})}{\left(\mathfrak R(V_t^{j,k}) \right)^2} \approx \dfrac{ -I(\mathfrak R(\dot{V}_t^{j,k,\mbox{\scriptsize normal}})+\mathfrak R(\dot{V}_t^{j,k,\mbox{\scriptsize motion}}))}{\left(\mathfrak R(V_t^{j,k}) \right)^2}}
\end{equation}
The approximation \eqref{transconductance_derivative} can be expressed as
\begin{equation}\label{conductance_motion}
	\dot{g}_t^{j,k} \approx \dot{g}_t^{j,k,\mbox{\scriptsize normal}} + \dot{g}_t^{j,k,\mbox{\scriptsize motion}}
\end{equation}
where 
\begin{equation}
	\dot{g}_t^{j,k,\mbox{\scriptsize normal}} = \dfrac{-I\mathfrak R(\dot{V}_t^{j,k,\mbox{\scriptsize normal}})}{(\mathfrak R(V_t^{j,k}))^2} \mbox{ and }
		\dot{g}_t^{j,k,\mbox{\scriptsize motion}} = \dfrac{-I\mathfrak R(\dot{V}_t^{j,k,\mbox{\scriptsize motion}})}{(\mathfrak R(V_t^{j,k}))^2}
\end{equation}
We note that, in the case of $\boldsymbol v_n=\boldsymbol 0$ in \eqref{VGMotion} (i.e., EIT measurement is not affected by motions), the relation \eqref{transconductance_derivative} becomes $\dot{g}_t^{j,k} = \dot{g}_t^{j,k,\mbox{\scriptsize normal}}$ by the reason of $V_t^{j,k,{\mbox{\scriptsize motion}}} = 0$. In the form of trans-conductance vector, the following approximation holds:
\begin{equation}
	\dot{\boldsymbol g}_t \approx \dot{\boldsymbol g}_t^{\mbox{\scriptsize normal}} + \dot{\boldsymbol g}_t^{\mbox{\scriptsize motion}}
\end{equation}
where
\begin{equation}
	\dot{\boldsymbol g}_t^{\mbox{\scriptsize normal}} = \begin{bmatrix} \dot{g}^{1,3,\mbox{\scriptsize normal}}_t, \cdots, \dot{g}^{16,14,\mbox{\scriptsize normal}}_t \end{bmatrix} \mbox{ and }\dot{\boldsymbol g}_t^{\mbox{\scriptsize motion}} = \begin{bmatrix} \dot{g}^{1,3,\mbox{\scriptsize motion}}_t, \cdots , \dot{g}^{16,14,\mbox{\scriptsize motion}}_t \end{bmatrix}
\end{equation}
If $\dot{\boldsymbol g}_t^{\mbox{\scriptsize normal}}$ satisfies the relation \eqref{noMAcondi}, we consequently obtain
\begin{equation}
	\dot{\boldsymbol g}_t \approx \dot{\boldsymbol g}_t^{\mbox{\scriptsize air}} + \dot{\boldsymbol g}_t^{\mbox{\scriptsize blood}} + \dot{\boldsymbol g}_t^{\mbox{\scriptsize motion}}
\end{equation}
Here, we note that $\dot{\boldsymbol g}_t^{\mbox{\scriptsize motion}}$ becomes more significant as motion (i.e., $|\boldsymbol v_n|$ in \eqref{VGMotion}) is large.

\section{Machine Learning Models} \label{appendixB}
\subsection{Discriminative Models}\label{appendixB1}
\subsubsection*{Logistic Regression (LR)} 
A LR model $f_{\mbox{\scriptsize LR}}$ consists of linear transformation and sigmoid as follows:
\begin{equation}
	f_{\mbox{\scriptsize LR}}(\overline{\boldsymbol X})=
	\sigma(\mathbf{w}^T\overline{\boldsymbol X}+b)
\end{equation}
where $\mathbf{w}\in\mathbb{R}^{150}$ and $b\in\mathbb{R}$ are learnable weight and bias, and $\sigma$ is a sigmoid function given by $\sigma(x)=(1+\exp(x))^{-1}$.

\subsubsection*{Multilayer Perceptron (MLP)} 
A MLP model $f_{\mbox{\scriptsize MLP}}$ has a hierarchical structure with nonlinearity compared to LR. Each layer consists of linear transformation and nonlinear activation. In our MLP models, ReLU is used in all layers except the last to avoid gradient vanishing \cite{Goodfellow2016}. Table \ref{Architecture} shows the architectures of the MLPs used in this study.

\subsubsection*{Convolutional Neural Network (CNN)} 
A CNN model $f_{\mbox{\scriptsize CNN}}$ consists of two paths; 1) feature extraction and 2) classification paths. In this study, the feature extraction path is based on VGG16 \cite{Simonyan2014}, as shown in Table \ref{Architecture}. The resultant feature map is flattened and then forwarded to the classification path, which is a MLP. 

The feature extraction path is a series of two convolutional and maxpooling (or flatten) layers, whose depth is associated with receptive field (RF) size of a unit in the last convolutional layer \cite{Luo2017}. According to the length of this series, VGG16-3, -4, and -5 are defined, where 3, 4, and 5 represent the iteration number of the layers in the series. Here, RFs are given by 32, 68, and 140, respectively.

\begin{table}[h]
	\centering
	{\scriptsize (a) MLP1 (MLP2) \\
		\begin{tabular}{|c|c|c|c|} \hline
			Layer & Input Dim & Output Dim & Activation \\ \hline
			Linear & 150 (150) & 150 (150) & ReLU \\ 
			Linear & 150 (150) & 300 (150) & ReLU \\
			Linear & 300 (150) & 300 (100) & ReLU \\
			Linear & 300 (100) & 150 (50) & ReLU \\
			Linear & 150 (50) & 150 (25) & ReLU \\
			Linear & 150 (25) & 150 (10) & ReLU \\ \hline
			Linear & 150 (10) & 1 (1) & Sigmoid \\ \hline
		\end{tabular} \\ \vspace{0.1cm}
		(b) VGG16-5; [1] Feature extraction and [2] Classification networks\\
		\begin{tabular}{|c|c|c|c|c|c|c|} \hline
			& Layer & Input Dim & Output Dim & Kernel & Activation & RF \\ \hline
			\multirow{18}{*}{[1]} & Conv1D & 150$\times$1 & 150$\times$4 & 3$\times$4 & ReLU & 3 \\ 
			& Conv1D & 150$\times$4 & 150$\times$4 & 3$\times$4 & ReLU & 5 \\ 
			& MaxPool1D & 150$\times$4 & 75$\times$4 & 2 & ReLU & 6 \\ \cline{2-7}
			& Conv1D & 75$\times$4 & 75$\times$8 & 3$\times$8 & ReLU & 10 \\ 
			& Conv1D & 75$\times$8 & 75$\times$8 & 3$\times$8 & ReLU & 14 \\ 
			& MaxPool1D & 75$\times$8 & 37$\times$8 & 2 & ReLU & 16 \\ \cline{2-7}
			& Conv1D & 37$\times$8 & 37$\times$16 & 3$\times$16 & ReLU & 24 \\ 
			& Conv1D & 37$\times$16 & 37$\times$16 & 3$\times$16 & ReLU & 32 \\ 
			& MaxPool1D & 37$\times$16 & 18$\times$16 & 2 & ReLU & 36 \\\cline{2-7}
			& Conv1D & 18$\times$16 & 18$\times$32 & 3$\times$32 & ReLU & 52 \\ 
			& Conv1D & 18$\times$32 & 18$\times$32 & 3$\times$32 & ReLU & 68 \\ 
			& MaxPool1D & 18$\times$32 & 9$\times$32 & 2 & ReLU & 76 \\ \cline{2-7}
			& Conv1D & 9$\times$32 & 9$\times$64 & 3$\times$64 & ReLU & 108 \\ 
			& Conv1D & 9$\times$64 & 9$\times$64 & 3$\times$64 & ReLU & 140 \\ 
			& Flatten & 9$\times$64 & 576$\times$1 & - & - & - \\ \hline
			\multirow{2}{*}{[2]} & Linear & 576$\times$1 & 576$\times$1 & - & ReLU & - \\ 
			& Linear & 576$\times$1 & 1$\times$1 & - & Sigmoid & -\\ \hline
	\end{tabular}}
	\caption{Network architectures; MLP and VGG16-6.}
	\label{Architecture}
\end{table}

\subsection{Manifold-learning Models}\label{appendixB2}
This subsection explains structures of an encoder $\mathcal E$ and a decoder $\mathcal D$ in \eqref{AELoss}, which were used for the manifold-learning approach described in Section \ref{sec:semi-sup}. The dimension of the latent vector $\boldsymbol z$ was constantly set as 10 in our experiments.

\subsubsection*{Principal Component Analysis (PCA)}
PCA learns principal vectors $\{\boldsymbol v_i \in \mathbb{R}^{150} \}_{i=1}^{10}$ in the following sense: For $i=1,\cdots,10$,
\begin{equation}
	\boldsymbol v_i  = \underset{\|\boldsymbol v\|=1}{\mbox{argmax}} ~  \|\mathfrak X_i \boldsymbol v\|_{2}^2 \mbox{ and } \mathfrak X_i = \mathfrak X_{i-1} - \boldsymbol v_{i-1} \boldsymbol v_{i-1}^T 
\end{equation}
where $\mathfrak X_1 := [ \overline{\boldsymbol X}_{\mbox{\scriptsize pos}}^{(1)}, \overline{\boldsymbol X}_{\mbox{\scriptsize pos}}^{(2)}, \cdots, \overline{\boldsymbol X}_{\mbox{\scriptsize pos}}^{(N_{\mbox{\scriptsize pos}})} ]^T$. For ease of explanation, $\mathfrak X_1$ is assumed to be zero-mean. An encoder $\mathcal E_{\mbox{\scriptsize pca}}$ and a decoder $\mathcal D_{\mbox{\scriptsize pca}}$ are given by
\begin{equation}
	\mathcal E_{\mbox{\scriptsize pca}}(\overline{\boldsymbol X}) = \boldsymbol z := \begin{bmatrix} \langle \overline{\boldsymbol X}, \boldsymbol v_1  \rangle, \cdots, \langle \overline{\boldsymbol X}, \boldsymbol v_{10} \rangle \end{bmatrix} \mbox{ and }
	\mathcal D_{\mbox{\scriptsize pca}}(\boldsymbol z) = \sum_{j=1}^{10} \boldsymbol z_i \boldsymbol v_i
\end{equation}
where $\boldsymbol z_i$ is $i$-th component of $\boldsymbol z$.

\begin{table}[h]
	\centering
	{\scriptsize
		(a) VAE \\
		\begin{tabular}{|c|c|c|c|} \hline
			\multicolumn{4}{|c|}{Encoder} \\ \hline
			Layer & Input Dim & Output Dim & Activation \\ \hline
			Linear & 150 & 125 & ReLU \\ 
			Linear & 125 & 75 & ReLU \\
			Linear & 75 & 50 & ReLU \\	
			Linear & 50 & 10$\times 2$ & - \\ 
			Sampling & 10$\times 2$& 10 & - \\ \hline
			\multicolumn{4}{|c|}{Decoder} \\ \hline
			Linear & 10 & 50 & ReLU \\
			Linear & 50 & 75 & ReLU \\
			Linear & 75 & 125 & ReLU \\
			Linear & 125 & 150 & - \\ \hline
		\end{tabular} \\ \vspace{0.1cm}
		(b) Convolutional VAE \\
		\begin{tabular}{|c|c|c|c|c|} \hline
			\multicolumn{5}{|c|}{Encoder} \\ \hline
			Layer & Input Dim & Output Dim & Kernel & Activation \\ \hline
			Conv1D & 150$\times$1 & 75$\times$8 & 3$\times$8 & ReLU \\
			Conv1D & 75$\times$8 & 38$\times$16 & 3$\times$16 & ReLU \\
			Conv1D & 38$\times$16 & 19$\times$24 & 3$\times$24 & ReLU \\
			Conv1D & 19$\times$24 & 10$\times$32 & 3$\times$32 & ReLU \\
			Flattening & 10$\times$32 & 320$\times$1 & - & - \\
			Linear & 320$\times$1 & 10$\times$2 & - & - \\
			Sampling & 10$\times$2 & 10$\times$1 & - & - \\ \hline
			\multicolumn{5}{|c|}{Decoder} \\ \hline
			Linear & 10$\times$1 & 320$\times$1 & - & - \\
			Reshaping & 320$\times$1 & 10$\times$32 & - & - \\
			DeConv1D & 10$\times$32 & 19$\times$24 & 3$\times$24 & ReLU \\
			DeConv1D & 19$\times$24 & 38$\times$16 & 3$\times$16 & ReLU \\
			DeConv1D & 38$\times$16 & 75$\times$8 & 3$\times$8  & ReLU \\
			DeConv1D & 75$\times$8 & 150$\times$8 & 3$\times$8 & ReLU \\
			Conv1D & 150$\times$8 & 150$\times$1 & 1$\times$1 & ReLU \\
			Linear & 150$\times$1 & 150$\times$1 & - & \\ \hline
		\end{tabular}}
	\caption{VAE network architectures.}
	\label{Architecture2}
\end{table}

\subsubsection*{Variational Auto-encoder (VAE)}
Table \ref{Architecture2} shows encoder-decoder models for VAE, whose network architecture is based on either MLP or CNN. In VAE, $\boldsymbol z$ is given by the following sampling procedure: $\boldsymbol z = \boldsymbol \mu + \boldsymbol \sigma \odot \boldsymbol z_{\mbox{\scriptsize noise}}$ and $\boldsymbol z_{\mbox{\scriptsize noise}} \sim \mathcal N(\boldsymbol 0, \boldsymbol I)$, where $\boldsymbol \mu$ and $\boldsymbol \sigma$ are substantial outputs generated by a neural network, $\odot$ is the element-wise product, and $\mathcal N(\boldsymbol 0, \boldsymbol I)$ is the normal distribution of mean $\boldsymbol 0$ and covariance $\boldsymbol I$. Here, $\boldsymbol 0$ is the zero vector and $\boldsymbol I$ is the identity matrix of $10 \times 10$. 

For VAE training, the following term is added to the loss function \eqref{AELoss}:
\begin{equation}\label{VAEloss}
	\mbox{KL}(\mathcal N(\boldsymbol \mu,\boldsymbol \Sigma) \| \mathcal N(\boldsymbol 0, \boldsymbol I)) = \dfrac{1}{2} \sum_{i=1}^{10} (\boldsymbol \mu_i^2+ \boldsymbol \sigma_i^2-\log \boldsymbol \sigma_i - 1)
\end{equation}
where $\mbox{KL}$ is Kullback-Leibler divergence and $\boldsymbol \Sigma$ is a $10\times 10$ diagonal matrix whose $(i,i)$ entry is $\boldsymbol \sigma_i$. This term enables VAE to learn dense and smooth latent space embedding in or near $\mathcal N(\boldsymbol 0, \boldsymbol I)$ \cite{Kingma2013,Seo2019,Yun2022}.

\subsubsection*{$\beta$-Variational Auto-encoder ($\beta$-VAE)} $\beta$-VAE differs with VAE in terms of loss function while sharing a model architecture. For some $\beta \in \mathbb{R}$, $\beta \times \mbox{KL}$ is added to the loss \eqref{AELoss} instead of \eqref{VAEloss} (i.e., VAE is the case of $\beta=1$). This simple weighting is known to be advantageous on disentangled representation learning of underlying factors \cite{Higgins2016}. We determined an optimal $\beta$ as the empirical best. Table \ref{optimalbeta} showed SQI performance variation about $\beta$ in the dataset where the scale and size normalization using linear interpolation were applied.\\
\begin{table}[h]
	{\scriptsize
		\begin{center}
			\begin{tabular}{|p{0.75cm}|p{1.5cm}|p{0.9cm}|p{0.9cm}|p{0.9cm}|p{0.9cm}|p{0.9cm}|p{0.9cm}|p{0.9cm}|p{0.9cm}|p{0.9cm}|p{0.9cm}|} 
				\hline 
				\multicolumn{2}{|c|}{} & \multicolumn{5}{c|}{\textbf{$\boldsymbol \beta$-VAE}} & \multicolumn{5}{c|}{\textbf{$\boldsymbol \beta$-CVAE}} \\ \hline
				\multicolumn{2}{|c|}{} & $\boldsymbol \beta = \frac{1}{3}$ & $\boldsymbol \beta=\frac{1}{2}$ & $\boldsymbol \beta=1$ & $\boldsymbol \beta=2$ & $\boldsymbol \beta=3$ & $\boldsymbol \beta = \frac{1}{3}$ & $\boldsymbol \beta=\frac{1}{2}$ & $\boldsymbol \beta=1$ & $\boldsymbol \beta=2$ & $\boldsymbol \beta=3$ \\ \hline
				\multirow{6}{*}{\textbf{Test}} & \textbf{Accuracy} & 0.9060 & 0.9092 & 0.9066 & 0.8951 & 0.9221 & 0.9208 & 0.9298 & 0.9292 & 0.9195 & 0.9189 \\ 
				& \textbf{PPV} & 0.9694 & 0.9680 & 0.9687 & 0.9682 & 0.9672 & 0.9750 & 0.9739 & 0.9688 & 0.9663 & 0.9663 \\ 
				& \textbf{NPV} & 0.6151 & 0.6282 & 0.6181 & 0.5802 & 0.6900 & 0.6617 & 0.7011 & 0.7100 & 0.6720 & 0.6693 \\ 
				& \textbf{Sensitivity} & 0.9203 & 0.9255 & 0.9218 & 0.9084 & 0.9441 & 0.9322 & 0.9441 & 0.9486 & 0.9397 & 0.9389 \\ 
				& \textbf{Specificity} & 0.8142 & 0.8047 & 0.8095 & 0.8095 & 0.7952 & 0.8476 & 0.8380 & 0.8047 & 0.7904 & 0.7904 \\
				& \textbf{AUC} & 0.9503 & 0.9439 & 0.9513 & 0.9426 & 0.9489 & 0.9531 & 0.9603 & 0.9528 & 0.9528 & 0.9471 \\ \hline
			\end{tabular}
	\end{center}}
	\caption{$\boldsymbol \beta$-VAE performance comparison about varying $\boldsymbol \beta$.}
	\label{optimalbeta}
\end{table}

\bibliographystyle{wileynum}
\bibliography{\jobname}

\end{document}